\documentclass[12pt,superscriptaddress, aps, prl, preprint]{revtex4-1}

\usepackage{graphicx}
\usepackage{amsmath}
\usepackage{amsthm}
\usepackage{amsfonts}
\usepackage{epstopdf}
\usepackage{booktabs}
\usepackage[usenames,dvipsnames,svgnames,table]{xcolor}
\usepackage{array}

\begin{document}

\author{Antonio Caretta}
\email{antonio.caretta@elettra.eu}
\affiliation{Elettra-Sincrotrone Trieste S.C.p.A. Strada Statale 14 - km 163.5 in AREA Science Park 34149 Basovizza, Trieste, Italy}

\author{Barbara Casarin}
\affiliation{Elettra-Sincrotrone Trieste S.C.p.A. Strada Statale 14 - km 163.5 in AREA Science Park 34149 Basovizza, Trieste, Italy}
\affiliation{Universit\`a degli Studi di Trieste, Via A. Valerio 2, 34127 Trieste, Italy}

\author{Paola Di Pietro}
\author{Andrea Perucchi}
\author{Stefano Lupi}
\affiliation{Elettra-Sincrotrone Trieste S.C.p.A. Strada Statale 14 - km 163.5 in AREA Science Park 34149 Basovizza, Trieste, Italy}

\author{Valeria Bragaglia}
\affiliation{Paul-Drude-Institut f{\"u}r Festk{\"o}rperelektronik, Hausvogteiplatz 5-7, 10117 Berlin, Germany}
\author{Raffaella Calarco}
\affiliation{Paul-Drude-Institut f{\"u}r Festk{\"o}rperelektronik, Hausvogteiplatz 5-7, 10117 Berlin, Germany}

\author{Felix Rolf Lutz Lange}
\affiliation{Institute of Physics, RWTH Aachen University, 52056 Aachen, Germany}
\author{Matthias Wuttig}
\affiliation{Institute of Physics, RWTH Aachen University, 52056 Aachen, Germany}

\author{Fulvio Parmigiani}
\affiliation{Elettra-Sincrotrone Trieste S.C.p.A. Strada Statale 14 - km 163.5 in AREA Science Park 34149 Basovizza, Trieste, Italy}
\affiliation{International Faculty, University of Cologne, 50937 Cologne, Germany}

\author{Marco Malvestuto}
\affiliation{Elettra-Sincrotrone Trieste S.C.p.A. Strada Statale 14 - km 163.5 in AREA Science Park 34149 Basovizza, Trieste, Italy}

\title{Interband characterization and electronic transport control of nanoscaled GeTe/Sb$_2$Te$_3$ superlattices}

\begin{abstract}
The extraordinary electronic and optical properties of the crystal-to-amorphous transition in phase-change materials led to important developments in memory applications. 
A promising outlook is offered by nanoscaling such phase-change structures.  
Following this research line, we study the interband optical transmission spectra of nanoscaled GeTe/Sb$_2$Te$_3$ chalcogenide superlattice films. 
We determine, for films with varying stacking sequence and growth methods, the density and scattering time of the free electrons, and the characteristics of the valence-to-conduction transition. 
It is found that the free electron density decreases with increasing GeTe content, for sub-layer thickness below $\sim$3~nm.  
A simple band model analysis suggests that GeTe and Sb$_2$Te$_3$ layers mix, forming a standard GeSbTe alloy buffer layer. 
We show that it is possible to control the electronic transport properties of the films by properly choosing the deposition layer thickness and we derive a model for arbitrary film stacks. 
\end{abstract}


\maketitle


\section{Introduction}
\noindent Phase-change materials (PCMs) constitute a class of semiconductors characterized by two allotrope phases, a crystalline and an amorphous one, having distinct physical properties \cite{Lencer11, deringer15}. 
Such materials, discovered in the late 60s \cite{Ovshinsky68}, are already used in rewritable optical discs, such as DVDs and Blu-ray discs. 
They are also very promising to realize fast, non-volatile electronic memories (PCRAM). 
Because of the variety of PCMs a rigorous chemical definition is absent, although design schemes for some ternary compounds are well established \cite{mottdavis71, lencer08, Lencer11}. 
A practical requirement for a PCM, as underlined in Ref.~\cite{lencer08}, is that the switching between the two phases must be  reversible, efficient and repeatable, due to obvious technological reasons. 
Benchmark PCMs are alloys composed of Ge, Sb and Te -- along the so called GeTe-Sb$_2$Te$_3$ pseudobinary line -- thus denoted as “GST” \cite{chen86,yamada91,wuttig07}. 
Although the potential for applications of PCMs spans from dynamic memories \cite{wuttig05} to display fabrication \cite{Hosseini14}, at present the limits for the use of GSTs are the speed and the power required to switch between the two phases, being respectively of approximately 100~ns and 100~$\mu$W \cite{lee-burger09}. 
Hence a huge effort is devoted to find new solutions, and nanoscale structures seem to offer significant advantages. \cite{Lankhorst05}. 
Following this trend, phase-change heterostructures had been produced, showing to function with improved performances \cite{simpson11}. 
Here we focus on chalcogenide superlattice (CSL) films made by high temperature deposition of alternating nm-size layers of GeTe and Sb$_2$Te$_3$ \cite{kooi15}. \\
Prior to investigate the properties of CSLs, we review the charge carrier transport \cite{Siegrist11,Zhang12} and the structural \cite{ovshinsky69, Kolobov04,akola07,Kolobov11,jostwuttig15} properties of GST compounds. 
The crystal phase of GSTs is characterized by a strong dependence of the resistivity upon annealing treatment \cite{Siegrist11}. 
In fact, by increasing the temperature, the system undergoes an insulator-metal transition (IMT) at 275$^\circ$C, due to disorder reduction. \cite{Siegrist11}
Density functional theory calculations showed that the insulating state is caused by the localization of charge carriers in vacancy-rich areas, and the transition to the metallic state happens when vacancies reconfigure into ordered layers \cite{Zhang12}. 
The presence of vacancy layers in the high temperature annealed phase of GSTs was proven by high-resolution TEM and electron diffraction \cite{kooi02}. \\
The lattice structure of the crystalline Ge-Sb-Te based CSL is similar, by growth design, to the high temperature GST phase. 
As mentioned before CSL are formed by high temperature deposition -- although lower than the IMT -- of GeTe and Sb$_2$Te$_3$ layers. 
One might consequently expect that the as grown structure is characterized by well separated layers of pure GeTe and Sb$_2$Te$_3$. 
Recent TEM and EXAFS experiments \cite{kooi15,casarin16} have shown that the structure is indeed layered, but GeTe is not actually pure since it is intercalated into a Sb$_2$Te$_3$ quintuple layer. 
Please note that the vacancy layers, when two-dimensionally extended, are labelled van der Waals gaps. \cite{oeckler12} 
The similarity between the high-temperature-annealed GSTs \cite{Siegrist11} and the CSLs, besides the structure, is corroborated by the measured resistivity of the latter that decreases with increasing temperature, suggesting a metallic type conductivity \cite{unpublished1}. 
Nonetheless, it is not obvious whether also the dielectric properties of CSLs are analogous to those of GST compounds \cite{lucovsky73,littlewood79_diel, yokota89, huang10}. 
For instance one of the peculiar characteristics of crystalline GST is the high value of the static dielectric constant, which is evidence of resonance bonding. \cite{shportko08} 
Measuring the dielectric function of CLS will provide useful information either about the structure, or about the bonding character of the compounds. 
The dielectric function of CSLs can be estimated by transmittance experiments, probing simultaneously the character of the interband transition and the density of free electrons. \\

\noindent In this article we compare the optical transmission behaviour of films grown by DC magnetron sputtering (MS), with those grown by molecular beam epitaxy (MBE) used as reference samples. 
Transmission experiments are performed on films with thickness below 100~nm, to guarantee high film quality, and at normal incidence, to simplify the interpretation of the results. 
Our analysis supports recent results showing that the GeTe layers are not isolated but intercalated within a Sb$_2$Te$_3$ layer \cite{kooi15, casarin16}. 
We explain how intermixing modifies the conductivity of the film. 
In addition we show that it is possible to control the electronic transport properties of the CSL not only by an annealing treatment, as in Ref. \cite{Siegrist11}, but also by means of a specific stack design. 
In fact, as we will see, the CSL conductivity depends on the layer sequence, giving a new design pathway for the control of the material properties. \\

\section{Experimental}
CSL samples are grown by MS and MBE on Sb-passivated Si(111) surfaces of intrinsic 500~$\mu$m-thick wafers above 210 $^\circ$C. 
MS samples are films of 15 repetitions of [GeTe($N$~nm)/Sb$_2$Te$_3$(3~nm)], where $N=1,2,3$. 
For the MBE samples we use the intermediate N=2 case. 
MS annealed (ANN) samples -- at 250~$^\circ$C for 30 minutes -- are distinguished from the MS as deposited (ASD) ones. 
Transmission experiments in the energy range 0.06--1.23~eV were collected at SISSI beamline on the Elettra storage ring \cite{lupi07} by a $Bruker~VERTEX~70v$  spectrometer. 
The fitting function is determined for multilayer thin films optics \cite{python_tmm}.

\section{Fitting model and results}
\begin{figure}
		\includegraphics[width=\columnwidth]{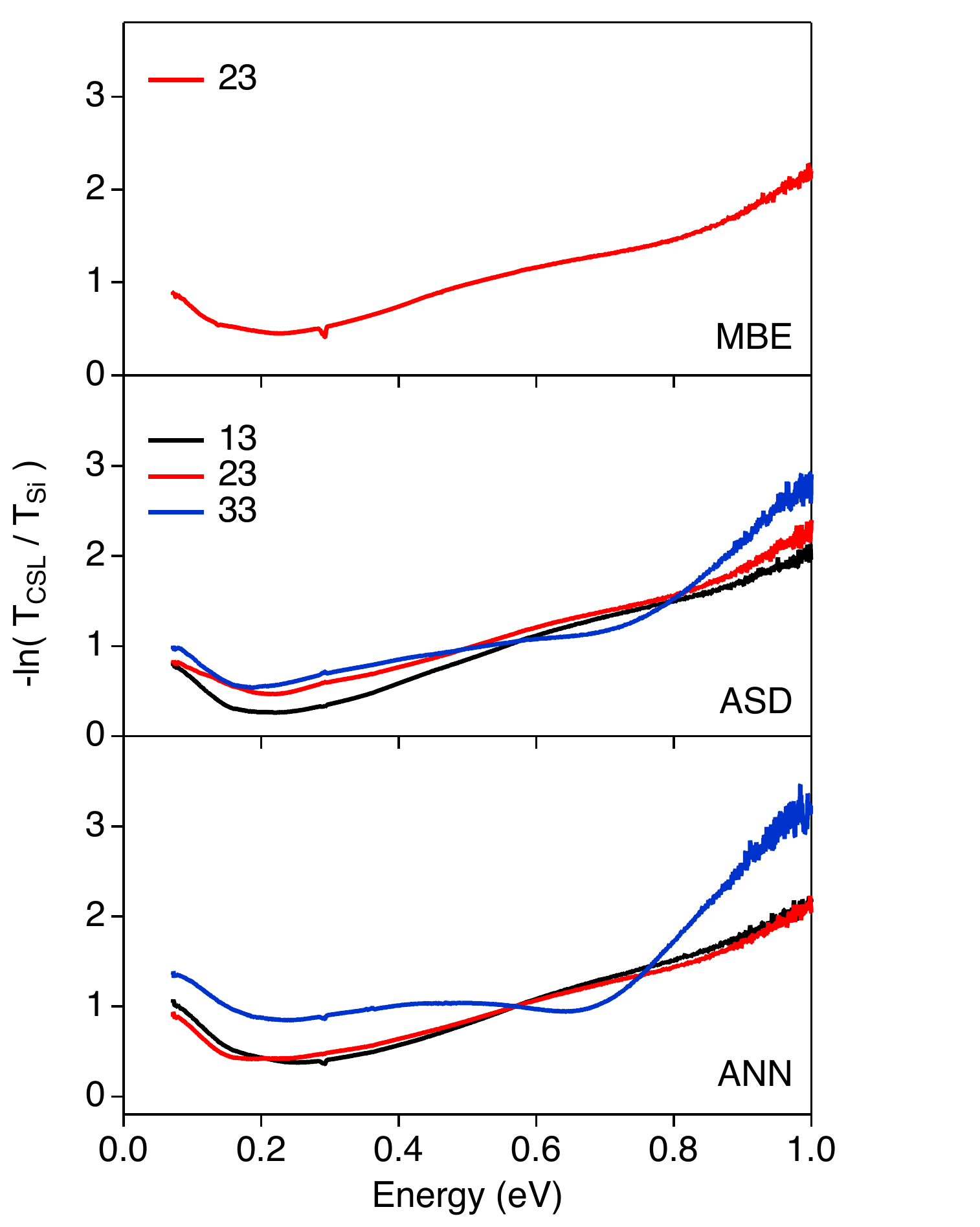}
	\caption{(Color online) Absorbance spectra of MBE 23, MS 13-23-33 as deposited (ASD) and MS 13-23-33 annealed (ANN) samples.} \label{tall}
\end{figure}

Figure \ref{tall} shows the absorbance $-$ln$(T_{CSL}/T_{Si})$, with $T_{CSL}$ and $T_{Si}$ respectively the sample and the substrate transmission, as a function of the incoming photon energy for all the samples. 
The data present few physically relevant features, common to all samples and regardless of growth technique and layer stack. 
In particular, we observe 
i) the onset of an absorption edge below $\sim$0.15~eV, increasing by decreasing the photon energy; 
ii) a minimum between 0.15 and 0.3~eV, and 
iii) a broad feature rising above $\sim$0.4~eV, increasing towards higher energies. 
In analogy to Ref.~\cite{shportko08} we associate the low frequency absorption component to free Drude electrons, and the higher energy broad band to the onset of the valence-to-conduction absorption. 
The absorption modulation of the data, particularly visible on the MS 33 annealed sample, is an artefact due to multilayer interference effects. 
We account for interference effects by calculating the transmission coefficient $T_{Data} = T_{CSL}/T_{Si}$ including reflection losses.\\

The film dielectric function (DF) is modelled with a sum of a Drude and a single Tauc-Lorentz (TL) oscillator term \cite{jellison96,shportko08}: 
\begin{equation} \label{eps}
\epsilon(x) = \epsilon^{Drude}(x) + \epsilon^{TL}(x),
\end{equation}
where 
$$\epsilon^{Drude}(x) = -\frac{\omega_p^2}{x^2+1/\tau^2} + i \frac{\omega_p^2}{x \tau (x^2+1/\tau^2)},$$ 
with $x$ the photon energy, $\omega_p$ the plasma frequency and $\tau$ the scattering time. 
The real part of $\epsilon^{Drude}$ is a negative Lorentzian function centred at zero while the imaginary part is positive and diverges at zero. 
The Tauc-Lorentz dielectric function is obtained by Kramers-Kronig integration ($\epsilon_1(x) = 1 + \frac{1}{\pi}\int_{-\infty}^{\infty} \frac{\epsilon_2(x')}{x'-x} dx'$) from the imaginary part:

\begin{equation} \label{epstl}
\left \{
\begin{array}{rcl}
\epsilon^{TL}_2 (x \leq E_g)	&=& 	0\\
\epsilon^{TL}_2 (x > E_g)	& = & 	\frac{A~C~E_0 ~(x-E_g)^2 }{ x~[(x^2-E_0^2)^2+C^2 x^2]}\\
\end{array}
\right. ,
\end{equation}

where $A$ is the TL amplitude, $E_g$ is the Tauc --or optical-- gap, 
$C$ is the band-width and $E_0$ is the central frequency, 
representing the photon energy where the transition probability is at the maximum. 
\begin{figure}
	\includegraphics[width=\columnwidth]{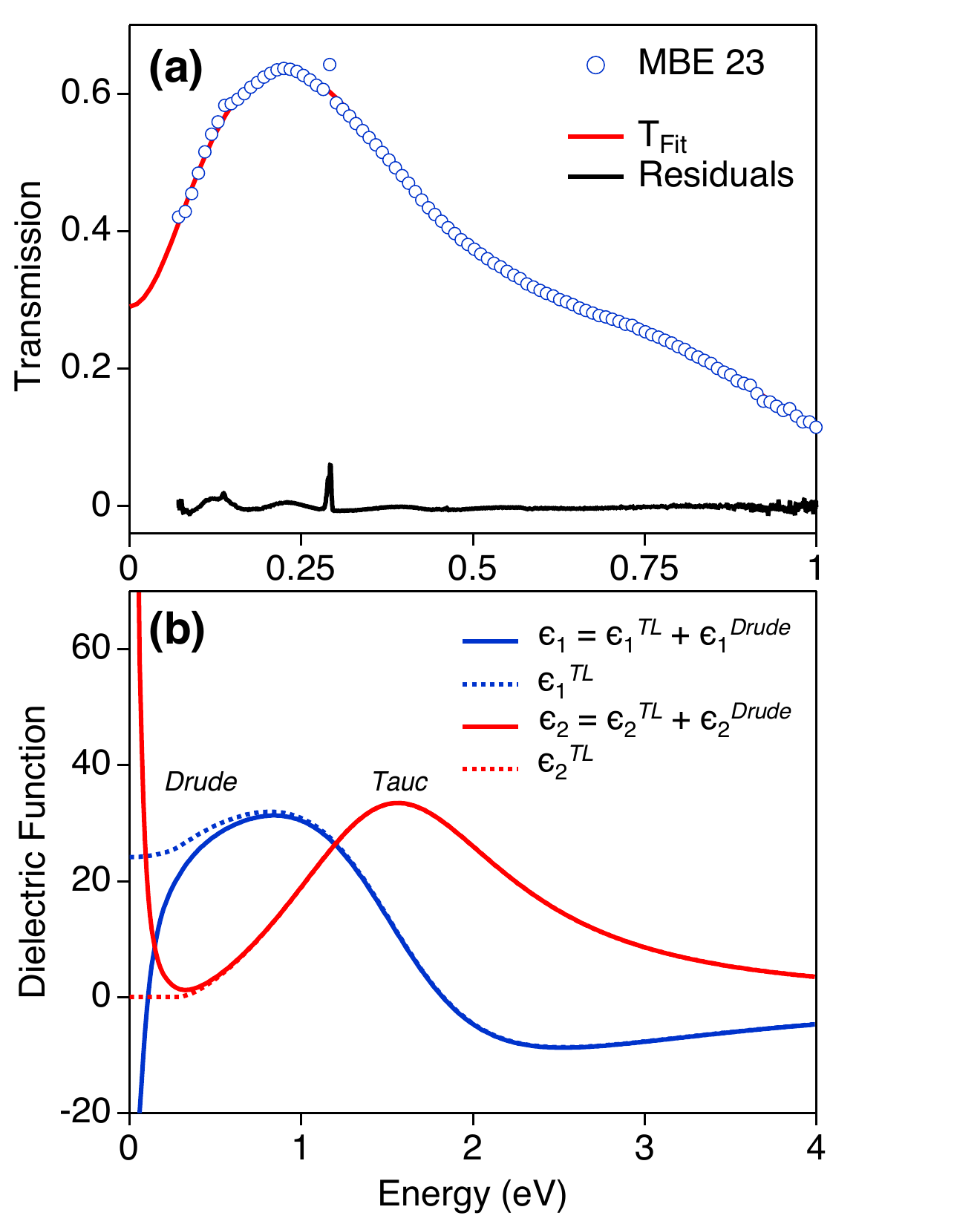}
	\caption{(Color online) a) Transmission data MBE~23 and fitting curve $T_{Fit}$. 
	 b) Dielectric function resulting from the fit. 
	 Both the real (red) and the imaginary (blue) components of the DF are shown also subtracting the Drude contribution (dashed).
	}\label{fitres}
\end{figure}
For the (intrinsic) silicon substrate we use $\epsilon(x) \sim 3.45$ \footnote{http://refractiveindex.info}. 
We verified the assumptions made for the dielectric properties of the substrate by transmittance measurements on virgin wafers. 
Regarding the calculation of $T_{Fit}(x)$ we include multiple coherent propagation only for the thin CSL film. 
Also the CSL film thickness is a free parameter of the fitting procedure. 
A typical fit result is shown in Figure \ref{fitres}a. 
The related DF is plotted in Figure \ref{fitres}b. 
The misfit between the experimental and calculated data is due to Si phonons (the peak at $\sim$0.3~eV) and probably to the small contribution of interference effect from the substrate that have not been accounted for.  
In all cases the fit function $T_{Fit}(x)$ describes all the main features of the transmission data, thus the deduced parameters are meaningful.\\
The fit parameters for the TL oscillator and the Drude term, for each investigated CSL, are given in Table \ref{tab_par}.\\
\begin{figure}
	\includegraphics[width=\columnwidth]{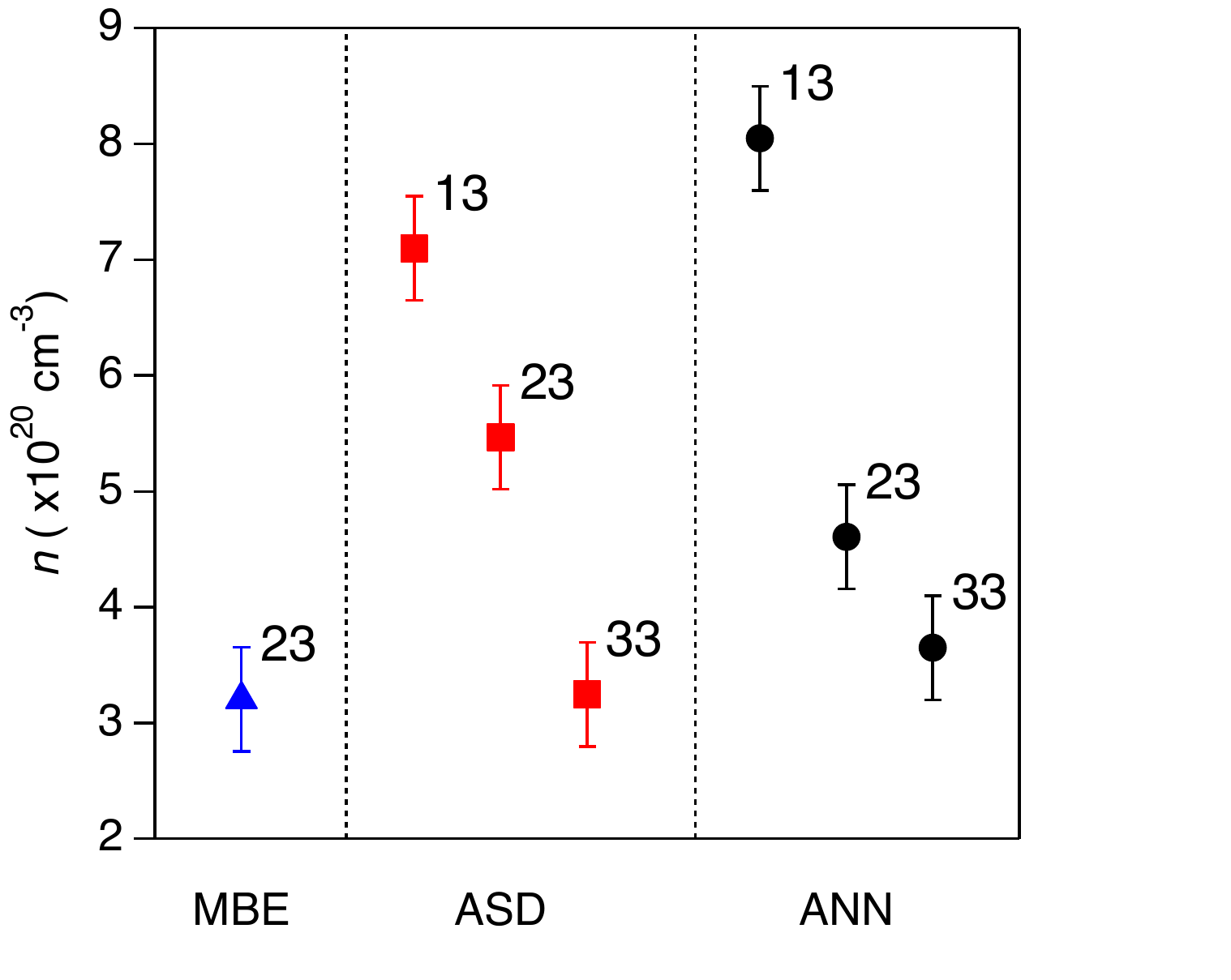}
	\caption{(Color online) Free electron density $n=\frac{m}{e^2} \omega_p^2$ obtained from the Drude term of the fit. 
	A decrease of $n$ is observed with increasing GeTe content.
	}\label{fin_par}
\end{figure}

\begin{table}\centering
\begin{ruledtabular}
\begin{tabular}{r c c c c c c}
 & \multicolumn{3}{c}{TL} & & \multicolumn{2}{c}{Drude}\\
 \hline
	& 	$E_g$(eV)  & $E_0$(eV) &  $\epsilon^{TL}_1(0)$ && $\omega_p$(eV) & $\tau$(fs) \\
\hline
MBE 23  &  0.3(1)	 & 1.7(2) & 23	&& 0.7(1) & 8(2) 	 \\
\hline
 13  	&  0.1(2) 	 & 1.5(2) & 35	&& 1.0(1) & 9(2) 	 \\
ASD 23  &  0.0(1) 	 & 1.5(1) & 29	&& 0.9(1) & 6(2) 	 \\
 33 	&  0.0(1) 	 & 1.5(1) & 24	&& 0.7(1) & 7(2) 	 \\
 \hline
 13   	&  0.1(2) 	& 1.3(1) & 30	&& 1.1(1) & 8(2) 	 \\
ANN 23  &  0.0(1) 	& 1.3(1) & 24	&& 0.8(1) & 13(2) 	 \\
 33   	&  0.5(1) 	& 1.5(1) & 23	&& 0.7(1) & 6(2) 	 \\
\end{tabular}
\end{ruledtabular}
	\caption{Paremeters resulting from the fit of Eq.~(\ref{eps}) for all the measured samples. 
	The error on the fit procedure is shown in parenthesis, representing the deviation of the last digit. 
	The low-energy limit of the TL DF, $\epsilon^{TL}_1(0)$, is also computed. }\label{tab_par}
\end{table}

\section{DISCUSSION}
We first consider the TL component of the DF, whose $E_g$(eV), $E_0$(eV),  $\epsilon^{TL}_1(0)$ parameters are given in Table \ref{tab_par}. 
The Tauc gap lays below 0.5~eV. 
$E_0$, representing the maximum of the semiconductor optical absorption, has an average value of approximately 1.5~eV. 
This value matches quite well with the experimentally obtained value of crystalline GeTe \cite{velea15} and GST \cite{shportko08}. 
Moreover, the computed 
low frequency value of the real part  of the TL DF, $\epsilon^{TL}_1(0)$, has an average value above 20. 
This $\epsilon^{TL}_1(0)$ high value is typical of the crystal phase of PCMs \cite{shportko08}, suggesting the presence of resonance bondings. 
In summary the TL component of our CSL is approximately independent on the annealing treatment. 
The fact that annealing does not affect the as-deposited (ASD) structures might suggest that the films undergo an immediate reorganization during growth, which is more effective than annealing itself. \cite{calarco16}
The analysis of the free electrons in our CSL supports this hypothesis.\\

\begin{figure}
	\includegraphics[width=0.7\columnwidth]{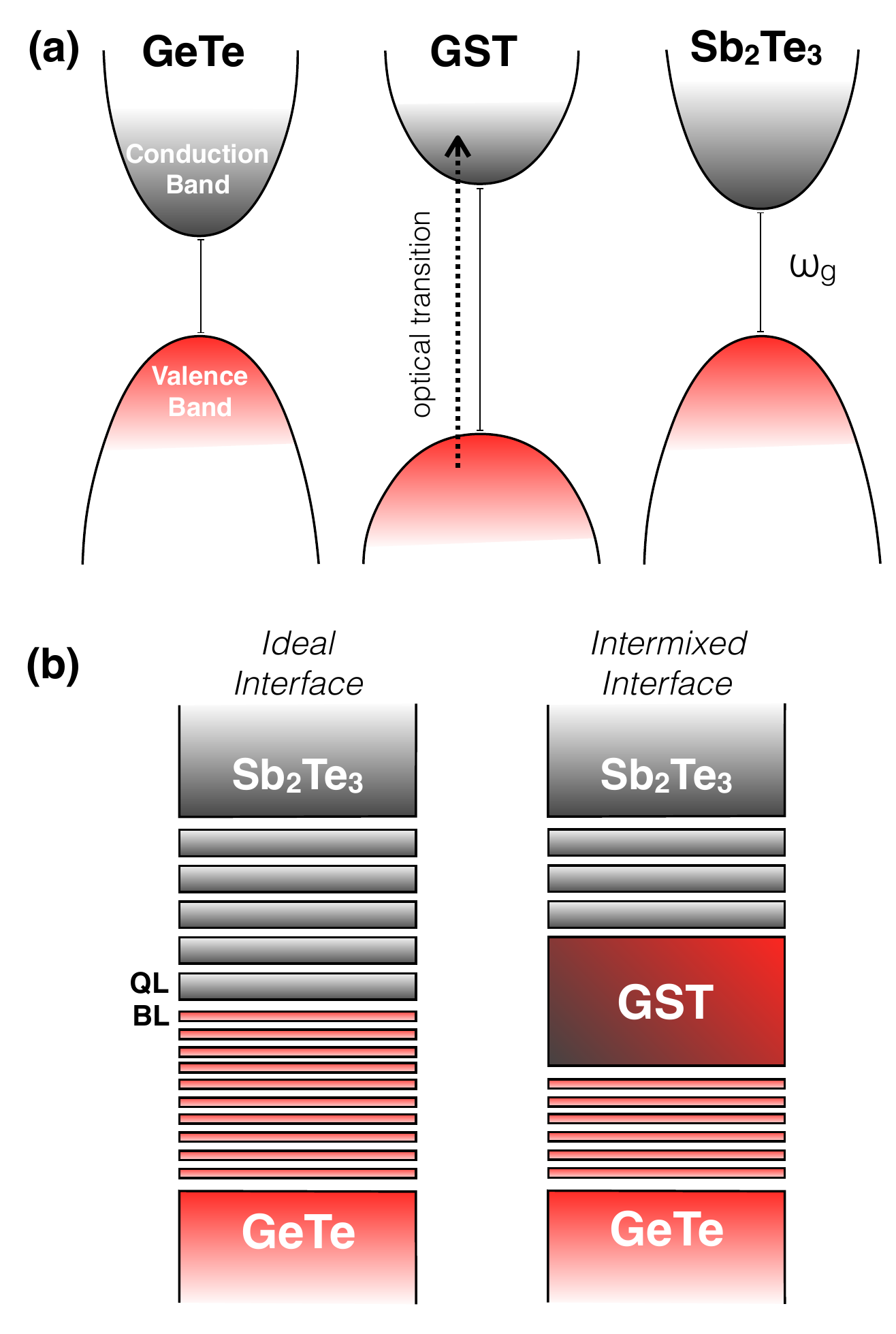}
	\caption{(Color online) a) Sketch of the band diagram of GeTe, Sb$_2$Te$_3$ and GST materials. 
The Tauc gap $E_g$ in both GeTe and Sb$_2$Te$_3$ is lower with respect to GST compounds. 
Thermal promotion of valence band (VB) electrons to the conduction band (CB) is accordingly reduced in GSTs, hence the free electron density and the conductivity are lower. 
b) Formation of GST layers at the interface between GeTe and Sb$_2$Te$_3$. } \label{scheme}
\end{figure}
The plasma frequency and the scattering time, resulting from the fitting procedure, are presented in Table \ref{tab_par}. 
While the scattering time ranges between 5-15 fs -- just half with respect to Copper at 300 K \cite{tanner} -- the plasma frequency consistently decreases with the increase of the stacking sequence 13-23-33, 
both for the ASD and ANN samples. 
The CSL electron density $n=\frac{m}{e^2} \omega_p^2$, derived from $\omega_p$, is shown in Figure \ref{fin_par}. 
A decrease of $n$ is observed with the increase of nominal GeTe content in the film. 
Also the conductivity $\sigma = \frac{e^2}{m} n \tau$ shows the same trend. 
This behaviour can be explained either by an increased defectivity, by a lower conductivity of GeTe with respect to Sb$_2$Te$_3$, or by an increase of the optical gap. 
Yet, since the comparison acts within the MS grown samples, we expect the same growth quality and -- consequently -- the same concentration of defects.  
In addition, pure GeTe has a \emph{higher} free carrier concentration with respect to Sb$_2$Te$_3$. \cite{GST95,tong15} 
Thus, the expected trend with stack sequence 13-23-33 should be opposite than what observed.
It follows that variations of $n$ must result from changes of the electronic band structure. \\

The intermixing of GeTe and Sb$_2$Te$_3$ layers rises the material optical gap, causing a reduction of the free carrier concentration. 
Figure \ref{scheme}a sketches the band diagrams for GeTe, Sb$_2$Te$_3$ and GSTs, in analogy to Refs \cite{ovshinsky69, Siegrist11}. 
The diagrams emphasize the differences between the optical gaps and the presence of degenerate electrons in the conduction band. 
Both GeTe and Sb$_2$Te$_3$ have a low Tauc gap $E_g \sim$0.1~eV \cite{esaki68, chopra69, ibram91}, while for GST $E_g^{GST} \sim$ 0.4~eV \cite{shportko08, kato05, park08}. 
In first approximation the Tauc gap of the CSL depends on the constituents volume fraction and on their respective gaps. 
If GeTe and Sb$_2$Te$_3$ are ideally separated inside the CSL the film is expected to have small gap and high carrier concentration, independent of the relative content of the two elements. 
If instead GST is formed by the intermixing of the two, the material Tauc gap must rise. 
The more GST is formed, the higher the Tauc gap (approaching eventually the value of GSTs) and the lower the number of degenerate electrons. 
This is consistent with the carrier density of the sequence 13-23-33, shown in Figure \ref{fin_par}. 

\section{Electric transport control}
\begin{figure}
	\includegraphics[width=0.8\columnwidth]{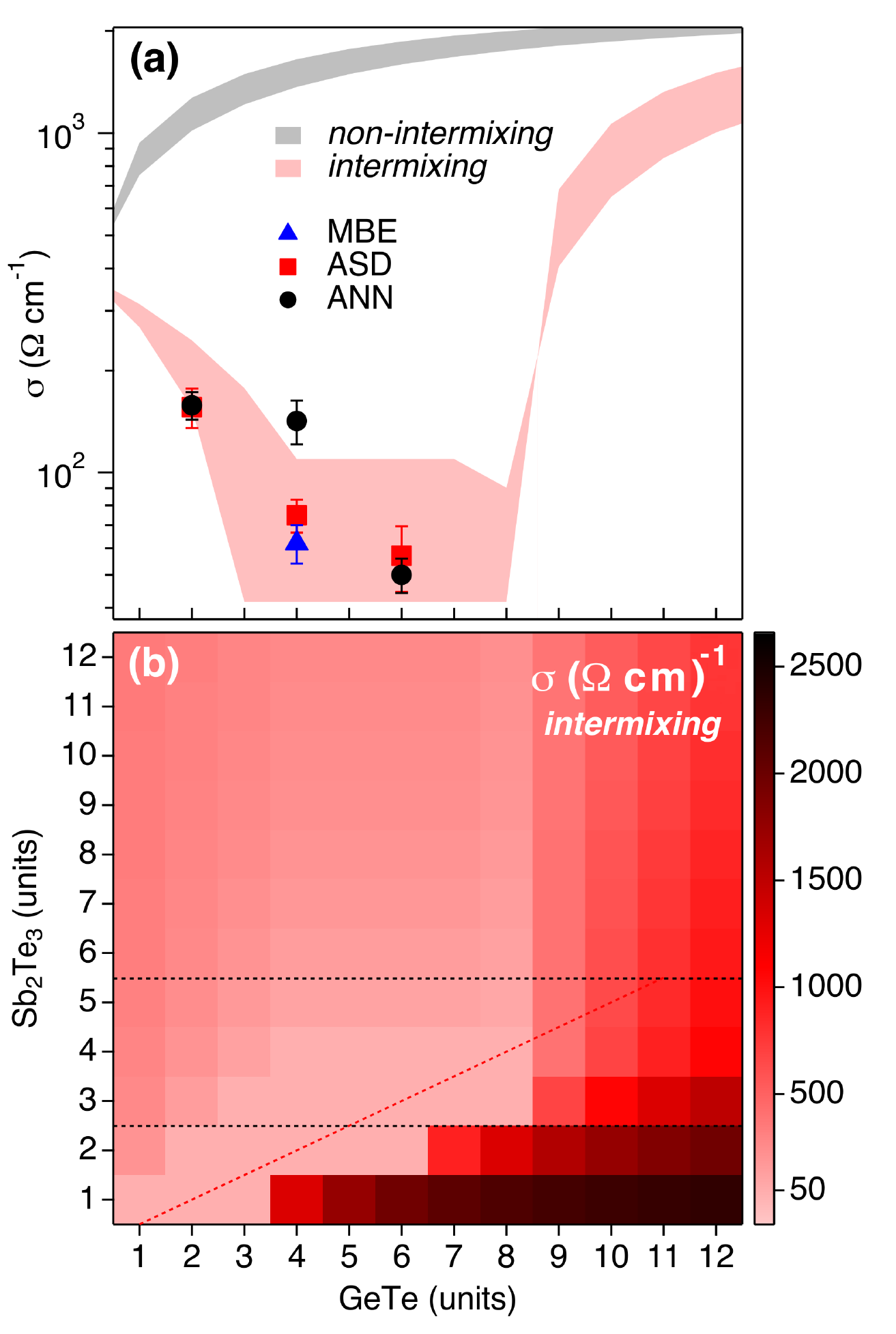}
	\caption{(Color online) a) Conductivity of the film calculated from Eq.~(\ref{sigma}) for increasing number of GeTe units, and compared with the measured conductivity of the CSLs. 
b) Conductivity of a film with stack sequence (GeTe)$_N$/(Sb$_2$Te$_3$)$_M$. 
The red dashed diagonal indicates the lowest conductivity region. 
The black dashed horizontal lines represent the cut used to generate the graph a).
}\label{NM}
\end{figure}

We extrapolate the trend observed in the present CSL to a stack of general thickness, having (GeTe)$_N$/(Sb$_2$Te$_3$)$_M$ as repeat unit, where $N,M \geq 1$. 
This formula is applicable to all CSL structures developed so far in literature: 
i) superlattice-like (SSL) PCMs, with $N,~M>2$ \cite{chong06}; 
ii) interfacial phase change materials (IPCM), with $N \leq 4,~M \leq 2$ \cite{simpson11}; 
iii) the case $N=2M$, studied in Ref.~\cite{tong15} 
and iv) our CSLs. 
We assume our stack sequence corresponds to $N=2, 4, 6$ and $M=3$, since 1~nm of GeTe along the (111) direction is composed by at least two bilayers \cite{Tominaga14,kooi15}. 
The conductivity of a CSL in the general case is calculated assuming that GeTe and Sb$_2$Te$_3$ layers intermix at the interface, forming GST as shown in Figure \ref{scheme}b. 
In particular, we consider that the interface layer consists of a maximum of 2 Sb$_2$Te$_3$ and 4 GeTe layers, resulting in 1 thick layer of GST. 
Then, given the bulk conductivities $\sigma_{GeTe}$, $\sigma_{Sb_2Te_3}$, $\sigma_{GST}$ \cite{tong15}, the film conductivity is calculated as:
\begin{equation}\label{sigma}
\sigma^{(N,M)} = \sum_{i} v_i \sigma_i
\end{equation}
where $v_i$ is the volume fraction of the $i$-th component (GeTe, Sb$_2$Te$_3$ and GST). \\
In Figure \ref{NM}a, the conductivity measured for all our samples is compared with the conductivity $\sigma^{(N,M=3)}$, as a function of GeTe layer units $N$. 
In agreement with experimental observations the conductivity decreases from $N=1$ down to $N \sim 8$. 
The strikingly different behaviour of the ``non-intermixing'' case, calculated simply via Eq.~(\ref{sigma}) with $v_{GST}=0$, demonstrates once more that intermixing effects must be accounted for when working with superlattice structures. \\
For completeness the dependence of $\sigma^{(N,M)}$ is shown in Figure \ref{NM}b. 
Interesting to note the conductivity has a minimum along the diagonal $M\sim N/2$ up to $N=8$ 
that corresponds to the maximum possible volumetric formation of GST, hence forming the material with the lowest conductivity. 
For higher values of $N$ and $M$ the conductivity approaches the ``non-intermixing'' case, as also visible for high values of GeTe units in Figure \ref{NM}a.\\

\section{Conclusion}
In this work we deduce important structural details of crystalline GeTe/Sb$_2$Te$_3$ superlattice deposited at high temperature ($>$210~$^\circ$C) by studying the film interband transmission. 
We observe that, at the interface between the GeTe and Sb$_2$Te$_3$ deposition layers, a crystalline GST compound is formed already during growth. 
GST, having higher band gap with respect to its constituents, lowers the number of degenerate conduction electrons. 
By varying the respective number of building block layers it is possible to control the percent of GST in the film, and consequently the film conductivity, with no need of annealing treatment. 
In the future it would be interesting to investigate the dynamics deriving from the intermixing and ion diffusion by depositing two single thick layers of GeTe and Sb$_2$Te$_3$ and studying the interface effects at different deposition and annealing conditions.

\section{Acknowledgements}
This work was supported by EU within FP7 project PASTRY [GA 317764]. 
AC would like to acknowledge Roberta Ciprian for useful comments.

%

\end{document}